\documentclass[aps,prl,twocolumn,superscriptaddress,nofootinbib]{revtex4-1}
\usepackage{amsmath}
\usepackage{amssymb}
\usepackage{graphicx}
\usepackage{braket}

\begin{document}

\title{Towards a room-temperature spin-photon interface based on nitrogen-vacancy centers and optomechanics}

\author{Roohollah Ghobadi}
\thanks{R. Ghobadi and S. Wein contributed equally to this work.\\
farid.ghobadi80@gmail.com; wein.stephen@gmail.com; christoph.simon@gmail.com}
\affiliation{Institute for Quantum Science and Technology and Department of Physics and Astronomy, University of Calgary, Calgary, Alberta, Canada T2N 1N4}
\affiliation{School of Physics, Institute for Research in Fundamental Sciences (IPM), Tehran, Iran}
\author{Stephen Wein}
\thanks{R. Ghobadi and S. Wein contributed equally to this work.\\
farid.ghobadi80@gmail.com; wein.stephen@gmail.com; christoph.simon@gmail.com}
\affiliation{Institute for Quantum Science and Technology and Department of Physics and Astronomy, University of Calgary, Calgary, Alberta, Canada T2N 1N4}
\author{ Hamidreza Kaviani}
\affiliation{Institute for Quantum Science and Technology and Department of Physics and Astronomy, University of Calgary, Calgary, Alberta, Canada T2N 1N4}
\author{Paul Barclay}
\affiliation{Institute for Quantum Science and Technology and Department of Physics and Astronomy, University of Calgary, Calgary, Alberta, Canada T2N 1N4}
\author{Christoph Simon}
\affiliation{Institute for Quantum Science and Technology and Department of Physics and Astronomy, University of Calgary, Calgary, Alberta, Canada T2N 1N4}

\date{\today}

\begin{abstract}
The implementation of quantum networks involving quantum memories and photonic channels without the need for cryogenics would be a major technological breakthrough. Nitrogen-vacancy centers have excellent spin properties even at room temperature, but phonon-induced broadening makes it challenging to coherently interface these spins with photons at non-cryogenic temperatures. Inspired by recent progress in achieving high mechanical quality factors, we propose that this challenge can be overcome using spin-optomechanical transduction. We quantify the coherence of the interface by calculating the indistinguishability and purity of single photons emitted from such a device and describe promising paths towards experimental implementation. Our results show that for ultra-high mechanical quality factor-frequency products, as have recently been achieved, our proposed interface could generate single photons with high indistinguishability, purity, and efficiency at room temperature---an important step towards room-temperature quantum networks.
\end{abstract}
\maketitle{}

Quantum networks will enable many applications such as secure communication \cite{QKD}, blind quantum computing \cite{Blind}, private database queries \cite{Private}, quantum clock networks \cite{clock}, more precise telescopes \cite{Telescope}, fundamental tests of quantum physics and quantum gravity \cite{Rideout}, and a future quantum internet \cite{SimonNP}. They require stationary qubits to store and process the quantum information and flying photonic qubits for communication. Interfacing stationary and flying qubits is therefore a critical component of any such network. Current implementations and proposals for quantum networks require cryogenic temperatures. Eliminating the requirement for cryogenics would be a major technological breakthrough.

Nitrogen-vacancy (NV) centers have excellent electron and nuclear spin properties even at room temperature \cite{T2NV,NuclearT2} and are thus promising candidates for stationary qubits. Entanglement between distant NV center spins has been created via the direct emission of photons at cryogenic temperatures \cite{Bernien}, enabling the first loophole-free test of Bell's inequality \cite{Hensen}. This approach cannot be implemented at room temperature because the NV center optical transition is dramatically broadened due to phonon interactions \cite{Fu}. As a consequence of the associated dephasing, emitted photons are not indistinguishable and cannot be used for entanglement generation via single-photon or two-photon interference \cite{SangouardRMP}.

Approaches to overcome this problem include looking for alternative stationary qubits that have good optical and spin properties at room temperature \cite{Newton, De Leon}, drastically speeding up the photon emission using ultra-small mode volume cavities \cite{Wein}, or a combination of both. We propose a promising strategy based on room-temperature quantum optomechanics, which has the additional advantage of allowing one to freely choose the photon wavelength. Thus the emission could be chosen to be in the telecom band, which is ideal for connecting distant stationary qubits through optical fibers.

The tremendous progress over the last few years in fabricating low-loss and low-mass mechanical oscillators and their capability to couple to a wide range of different systems, including light \cite{RMPom} and NV centers \cite{NVM1,NVM2}, make them attractive as potential quantum transducers \cite{Stannigel12, Bagci14, Lehnert}. Optomechanical systems are also emerging as an alternative for engineering light-matter interaction at the quantum level \cite{Verhagen,Wollman, Riedinger16,Galland} and for creating nonclassical correlations \cite{Hong17} even at room temperature \cite{purdy,sudhir}.

Experiments in the field of quantum optomechanics are usually done at cryogenic temperature to reduce detrimental thermal effects. However, very recent advances in achieving high mechanical quality factors in silicon-nitride (SiN) devices with ultra-high $Q_\text{m} f_\text{m}$ products \cite{OM roomT1,OM roomT2,OM roomT3,Ghadimi17} greatly facilitates room-temperature operation, allowing for our present proposal to use a spin-optomechanical system as a room-temperature spin-photon interface that emits single indistinguishable photons. 

\begin{figure}
\includegraphics[width= 0.8\linewidth]{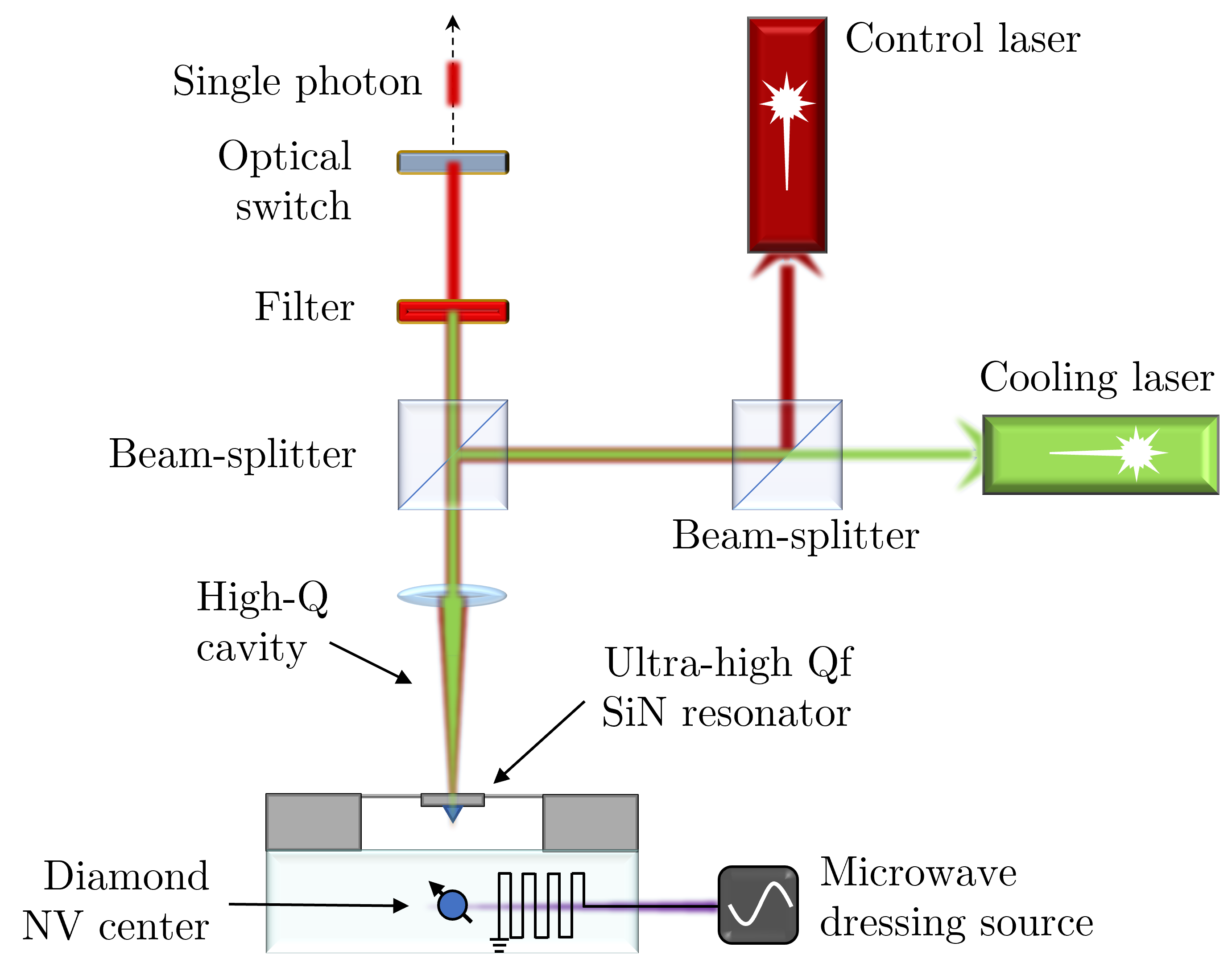}
\caption{Schematic of the proposed spin-optomechanical transducer emitting indistinguishable single photons. The optomechanical system consists of a trampoline membrane oscillator coupled to a high finesse optical cavity. A magnetic tip on the membrane couples the oscillator to the dressed ground states of an NV center implanted in a diamond substrate. The cooling laser keeps the oscillator near its ground state while the control laser induces the spin-photon interaction required to extract a single telecom photon.}
\label{diagram}
\end{figure}

Our approach is guided by previous work on spin-mechanical \cite{Rabl09} and microwave spin-optomechanical interactions \cite{Hybrid15, Li16}, which allows for engineering an effective Jaynes-Cumming interaction between dressed spin states and a photon. Figure \ref{diagram} shows a schematic of our proposal where a dressed NV center ground-state microwave spin excitation can be converted into a single photon in the telecom band. The spin-photon interaction is mediated by a mechanical oscillator that is controlled and cooled by two continuous-wave lasers.

The dressed spin system provides qubit states that can be robust against nuclear spin-bath fluctuations while also allowing for a strong tuneable microwave spin transition to facilitate a magnetic spin-mechanics coupling to the optomechanical device \cite{Rabl09,Dressed state}. The mechanical oscillator with resonance frequency $\omega_\text{m}=2\pi f_\text{m}$ is coupled to a Fabry-P\'{e}rot cavity by a radiation pressure interaction \cite{RMPom}. To generate a single photon (sp), a beam-splitter type optomechanical interaction is induced by driving a cavity mode ($\omega_\text{sp}$) with a red-detuned control laser ($\omega_\text{sp}^\text{L}$) near the phonon sideband ($\omega_\text{sp}-\omega_\text{m}$).

By adjusting the microwave dressing, the spin transition ($\omega_\text{q}$) can be tuned to match the detuning of the optomechanical control laser $(\omega_\text{q} = \omega_\text{sp}-\omega_\text{sp}^\text{L})$. When both the spin transition and the laser are detuned relative to the phonon sideband by $\delta = \omega_\text{q}-\omega_\text{m}$, the excited spin state will undergo a Raman transition to generate a single photon at the cavity frequency while only virtually exciting the mechanical oscillator \cite{Li16,Hybrid15}. From an energy conservation perspective, this process combines a spin excitation with a single control laser photon to produce a single photon at the cavity frequency. This Raman approach protects the spin from decoherence due to thermal fluctuations of the mechanical oscillator, allowing highly-indistinguishable emission.

To minimize the impact of thermal fluctuations and bring operation to room-temperature, the oscillator must be cooled during the spin-photon interaction. This can be achieved by optomechanical ground-state cooling \cite{RMPom} using the same interaction that mediates the single-photon extraction. However, the laser used to engineer a Raman transition is detuned from the phonon sideband by $\delta$ and does not provide efficient cooling. Also, using the same cavity mode for both single-photon extraction and cooling increases thermal noise in the single-photon channel. We propose to solve these problems by cooling (c) the oscillator using a different cavity mode ($\omega_\text{c}$) by driving the phonon sideband with a cooling laser ($\omega_\text{c}^\text{L}=\omega_\text{c}-\omega_\text{m}$).

Even with efficient cooling, the remaining thermal fluctuations still cause some constant thermal emission into the single-photon channel. To reduce this effect, and improve the performance of the interface, we also consider an optical switch at the output of the device to temporally post-select the single-photon.


We quantify the performance of the interface by constructing a numerical model and calculating the efficiency, indistinguishability \cite{Indistin1,Indistin2}, and purity \cite{Indistin1} of the single photon. We then discuss potential implementations and show that high performance is a realistic goal.

We model the two Fabry-P\'{e}rot cavity modes corresponding to $\omega_\text{sp}$ and $\omega_\text{c}$ with damping rates $\kappa_\text{sp}$ and $\kappa_\text{c}$ for single-photon extraction and optomechanical cooling, respectively. One mirror of the cavity serves as the mechanical oscillator with a damping rate of $\gamma_\text{m}$ and a mean phonon number $n_\text{th}=k_\text{B}T/\hbar\omega_\text{m}$. The spin-mechanics coupling to the nearby NV center is induced by a small magnetic tip attached to the oscillator.

The NV center spin triplet ground state has a zero-field splitting of $\omega_0/2\pi=2.88$ GHz separating the $\ket{m_s=0}$ and $\ket{m_s=\pm 1}$ states, a spin relaxation rate of $\gamma$, and a decoherence rate of $\gamma^\star\gg \gamma$. For an NV center with its z-axis aligned with the magnetic field, the NV-mechanics coupling is $\lambda(\hat{b}+\hat{b}^{\dagger})\hat{S}_{z}$, where $\hat{b}$ is the phonon annihilation operator, $\hat{S}_{z}$ is the $z$ component of the spin, $\lambda=2\mu_{B}x_\text{zpf}G_\text{m}/\hbar$ is the spin-mechanics coupling rate, $x_\text{zpf}=\sqrt{\hbar/2m\omega_\text{m}}$ is the zero-point fluctuation amplitude with an effective mass $m$, and magnetic field gradient $G_\text{m}$.

\begin{figure}
\includegraphics[width=0.9 \linewidth]{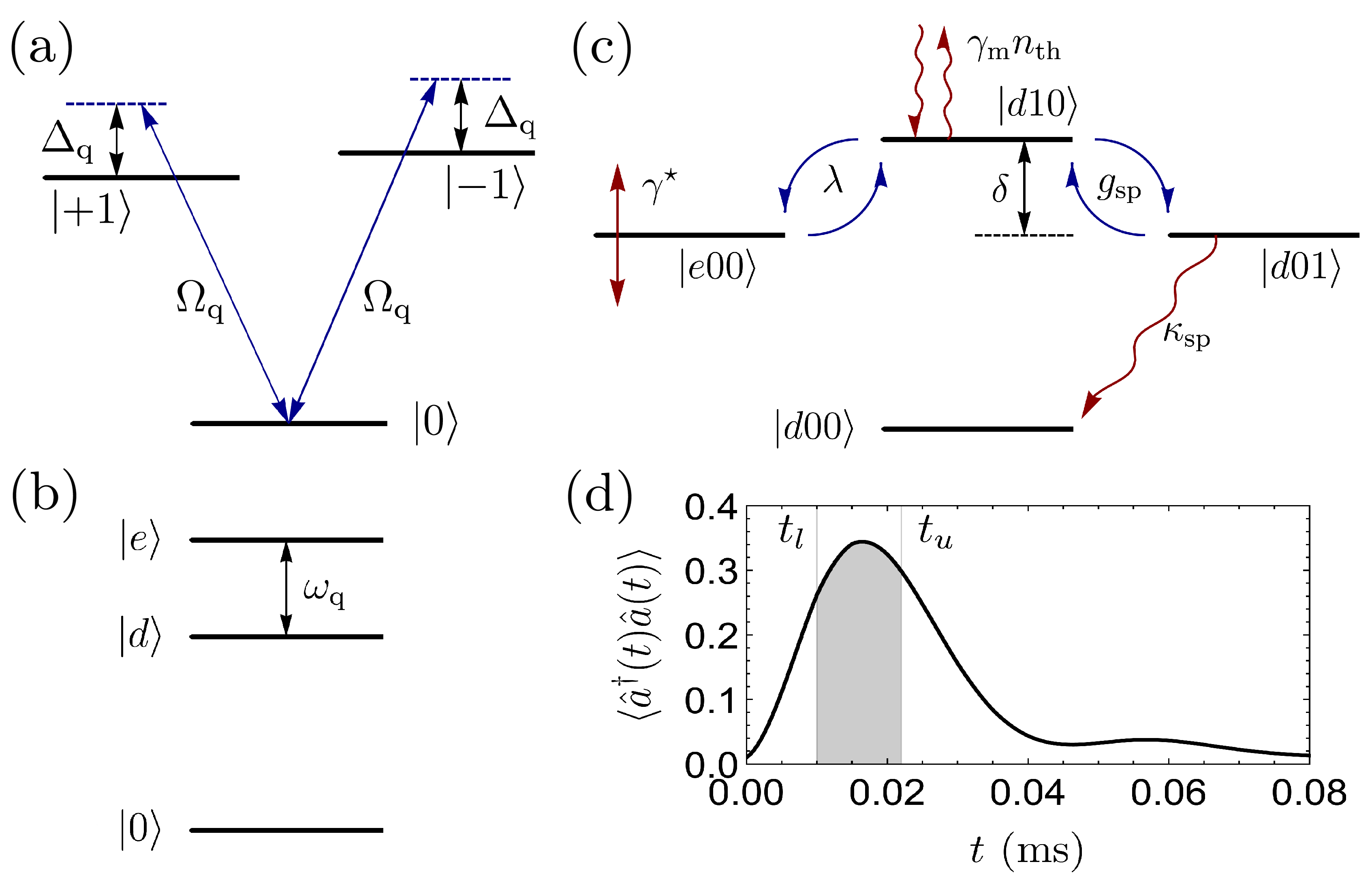}\vspace{-4mm}
\caption{(a) The NV ground-state spin triplet dressed by a microwave field with Rabi frequency $\Omega_\text{q}$ and detuning $\Delta_\text{q}$. (b) The dressed states $\ket{e,d}=(\ket{+1}\pm\ket{-1})/\sqrt{2}$ provide a tunable system with $\omega_\text{q}=\Omega_\text{q}^2/\Delta_\text{q}$ that couples to the mechanical oscillator. (c) The level diagram of the effective Raman transition. The excited spin state $\ket{e00}$ with decoherence rate $\gamma^\star$ is coupled to the $\delta$-detuned state $\hat{b}^\dagger\!\ket{d00}\!=\!\ket{d10}$ by magnetic coupling rate $\lambda$; $\ket{d10}$ then couples to $\hat{a}^\dagger\!\ket{d00}\!=\!\ket{d01}$ by the optomechancial coupling rate $g_\text{sp}$. This indicates the absorption of a control laser photon and oscillator phonon to generate a single photon that is released by the cavity at the rate $\kappa_\text{sp}$, leaving the system in $\ket{d00}$. In the adiabatic regime, $\ket{d10}$ is bypassed due to the Raman detuning $\delta$, allowing the effective coupling rate $\Omega=\lambda g_\text{sp}/\delta$. (d) An illustration of the single-photon wavepacket with lower and upper gate times.
}
\label{setup}
\end{figure}

To create a tunable transition for spin-mechanics coupling, we consider NV spin states dressed by a microwave field with Rabi frequency $\Omega_\text{q}$ and detuning $\Delta_\text{q}\gg \Omega_\text{q}$ \cite{Rabl09} (see Fig. \ref{setup}). For splittings on the order of $\omega_\text{q}\simeq 2\pi$ MHz $\ll\omega_0$, the NV center's Hamiltonian becomes $\hat{H}_\text{NV}=\omega_\text{q}\hat{\sigma}_+\hat{\sigma}_-$ where $\omega_\text{q}=\Omega_\text{q}^2/\Delta_\text{q}$ and $\hat{\sigma}_-=\ket{d}\bra{e}$ is the lowering operator for qubit states  $\ket{e}=(\ket{-1}+\ket{+1})/\sqrt{2}$ and $\ket{d}=(\ket{-1}-\ket{+1})/\sqrt{2})$. Then the spin-mechanics coupling becomes $\lambda(\hat{\sigma}_-+\hat{\sigma}_+)(\hat{b}+\hat{b}^\dagger)$.

The total system Hamiltonian, after linearizing the optomechanical interactions \cite{RMPom}, is given by \mbox{$\hat{H} = \omega_\text{q}(\hat{\sigma}_+\hat{\sigma}_-+\hat{a}^\dagger\hat{a})+\omega_\text{m}(\hat{b}^\dagger\hat{b}+\hat{c}^\dagger\hat{c})+\hat{H}_\text{I}$} with an interaction term \mbox{$\hat{H}_\text{I}=\left(\lambda\hat{\sigma}_-+g_\text{sp}\hat{a}+g_\text{c}\hat{c}\right)(\hat{b}+\hat{b}^\dagger)+\text{h.c.}$}, where $\hat{a}$ and $\hat{c}$ are the cavity mode annihilation operators for the single-photon and cooling modes, respectively. The rates $g_\text{sp}$ and $g_\text{c}$ are the optomechanical coupling rates, which can be tuned by adjusting the respective intensities of the driving lasers. We model the total dynamics of the system using the master equation

\begin{equation}
\begin{split}
\dot{\rho}=-i[\hat{H},\hat{\rho}]+\gamma^\star{\cal D}[\hat{\sigma}_+\hat{\sigma}_-]\hat{\rho}+\kappa_\text{sp}{\cal D}[\hat{a}]\hat{\rho}+\kappa_\text{c}{\cal D}[\hat{c}]\hat{\rho} \\
+\gamma_\text{m}n_\text{th}{\cal D}[\hat{b}^\dagger]\hat{\rho}+\gamma_\text{m}(1+n_\text{th}){\cal D}[\hat{b}]\hat{\rho},
\end{split}
\label{roeALL}
\end{equation}
where ${\cal D}[\hat{A}]\hat{\rho}=\hat{A}\hat{\rho}\hat{ A}^{\dagger}-\hat{A}^{\dagger}\hat{A}\hat{\rho}/2-\hat{\rho} \hat{A}^{\dagger}\hat{A}/2$. Figure \ref{setup}c shows a reduced level diagram of the system illustrating the effective Raman transition.

We model the optical switch using a rectangular gate function where $t_l$ is the beginning time and $t_u$ is the end time of the gate. An illustration of the emitted wavepacket and gate times is shown in figure \ref{setup}d. The brightness of this wavepacket is defined as the average number of emitted photons: $\beta=\kappa_\text{sp}\int_{t_l}^{t_u}dt\braket{\hat{a}^\dagger(t)\hat{a}(t)}$. This brightness closely approximates the single-photon efficiency when the purity is high. We characterize the purity by the integrated second-order correlation function \mbox{$g^{(2)}\!=\!\int_{t_l}^{t_u}dt\int_{0}^{t_u-t}\!d\tau
\braket{\hat{a}^\dagger(t)\hat{a}^\dagger(t+\tau)\hat{a}(t+\tau)\hat{a}(t)}/N$}, where \mbox{$N=
\int_{t_l}^{t_u}dt\int_{0}^{t_u-t}d\tau \braket{\hat{a}^\dagger(t)\hat{a}(t)}\braket{\hat{a}^\dagger(t+\tau)\hat{a}(t+\tau)}$}. A small $g^{(2)}$ corresponds to high single-photon purity. Similarly, we define the indistinguishability as the probability for two source photons to interfere \mbox{$I=\int_{t_l}^{t_u}dt\int_{0}^{t_u-t}d\tau
|\braket{\hat{a}^\dagger(t+\tau)\hat{a}(t)}|^2/N$} \cite{Indistin1,Indistin2}.

For sufficient optomechanical cooling $\braket{\hat{b}^\dagger\hat{b}}\ll 1$, a large Raman detuning $\delta\gg \lambda, g_\text{sp}$, and when the rotating-wave approximation is valid ($\delta,g_\text{c}\ll \omega_\text{q}, \omega_\text{m}$), the mechanical oscillator can be adiabatically eliminated to obtain an effective spin-photon coupling \cite{Li16,Hybrid15}. Reaching this regime is difficult at room temperature; however, a coherent spin-photon interface can still be achieved by only weakly satisfying these requirements. We verify this by numerically solving the full system described by equation (\ref{roeALL}), taking into account both single-photon extraction and optomechanical cooling without adiabatic elimination or rotating-wave approximations.

In the adiabatic regime, a Jaynes-Cummings type interaction dominates $\Omega(\hat{a} \hat{\sigma}_{+}+\hat{a}^{\dagger}\hat{\sigma}_{-})$, where \mbox{$\Omega=g_\text{sp}\lambda/\delta$} and the effective thermal noise becomes $\Gamma_\text{th}=g_\text{sp}\lambda n_\text{th}\gamma_\text{m}/\delta^2$ \cite{Li16}. The advantage of this regime becomes apparent by increasing $\delta$ until the coupling rate exceeds the thermal noise. In this regime, the single-photon generation is mathematically similar to a cavity-emitter Purcell effect in the critical regime \cite{Wein}. We follow methods similar to \cite{Indistin2,Wein} to identify that the effective rate of single-photon emission is $R=4\Omega^{2}/(\kappa_\text{sp}+2\gamma^\star+4\Gamma_\text{th})$ and the maximum single-photon efficiency is $\beta_0=R/(R+2\Gamma_\text{th})$. Therefore a highly efficient interface requires $R\gg\Gamma_\text{th}$. In addition, the indistinguishability is maximized near the critical coupling condition $R=\kappa_\text{sp}$ \cite{Wein}, which for $\kappa_\text{sp}\gg\gamma^\star,\Gamma_\text{th}$ implies $\delta\kappa_\text{sp}=2g_\text{sp}\lambda$.

To simulate the performance, we first allow the system to reach an equilibrium state under influence of the cooling laser. Then we prepare the spin state in $\ket{e}$ to compute $\hat{a}(t)$ and its associated correlation functions using a numerical implementation of quantum regression theorem with Runge-Kutta integration methods. 

\begin{figure}
\begin{flushleft}
\mbox{(a)\hspace{40mm}(b)}
\end{flushleft}\vspace{-5mm}

\hspace{-5mm}
\includegraphics[scale=0.4]{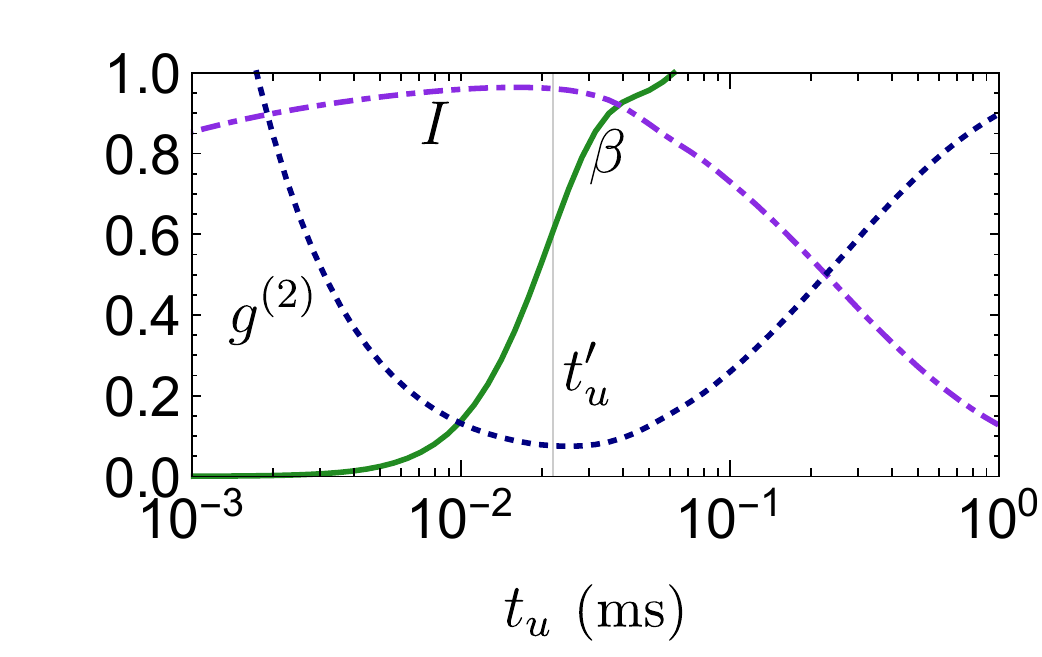}\hspace{-4mm}
\includegraphics[scale=0.4]{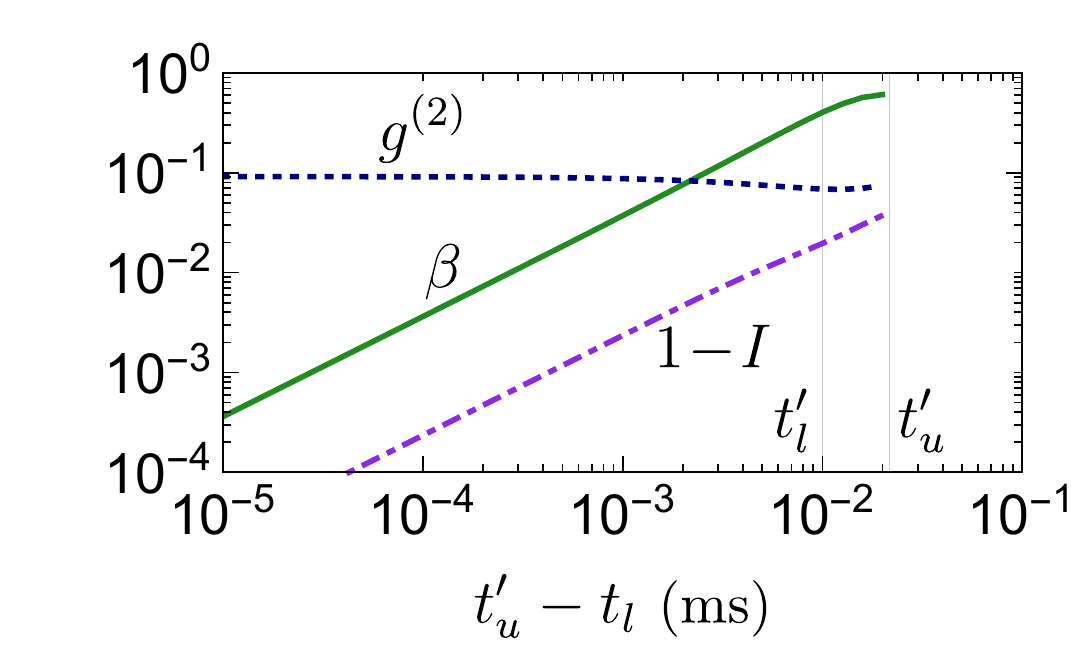}\vspace{-5mm}

\vspace{-3mm}
\begin{flushleft}
\mbox{(c)\hspace{40mm}(d)}
\end{flushleft}\vspace{-5mm}

\hspace{-5mm}
\includegraphics[scale=0.4]{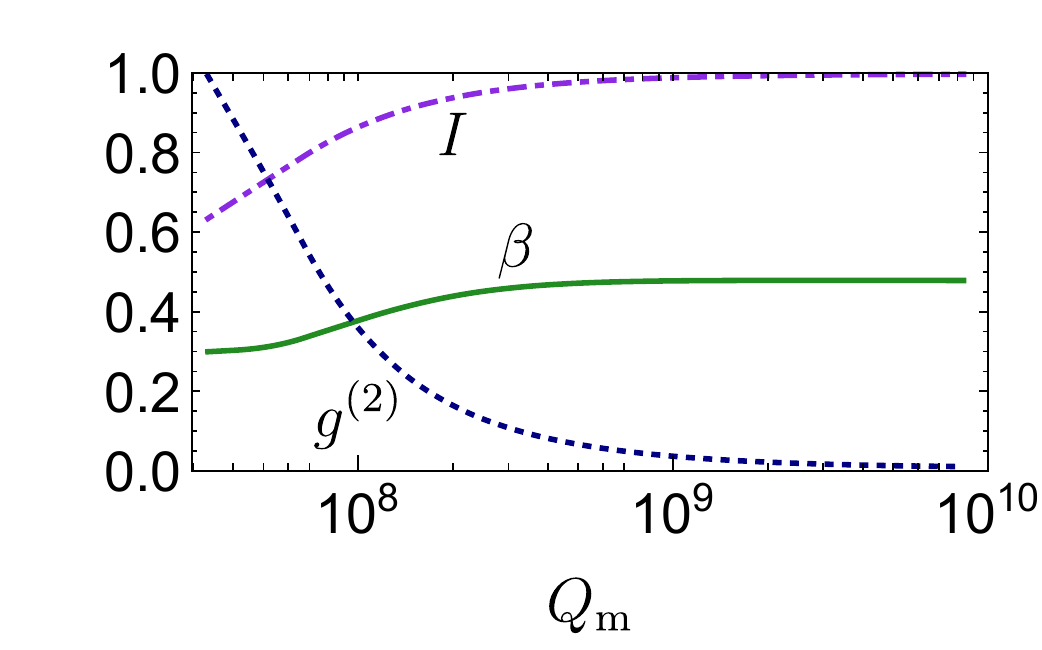}\hspace{-4mm}
\includegraphics[scale=0.4]{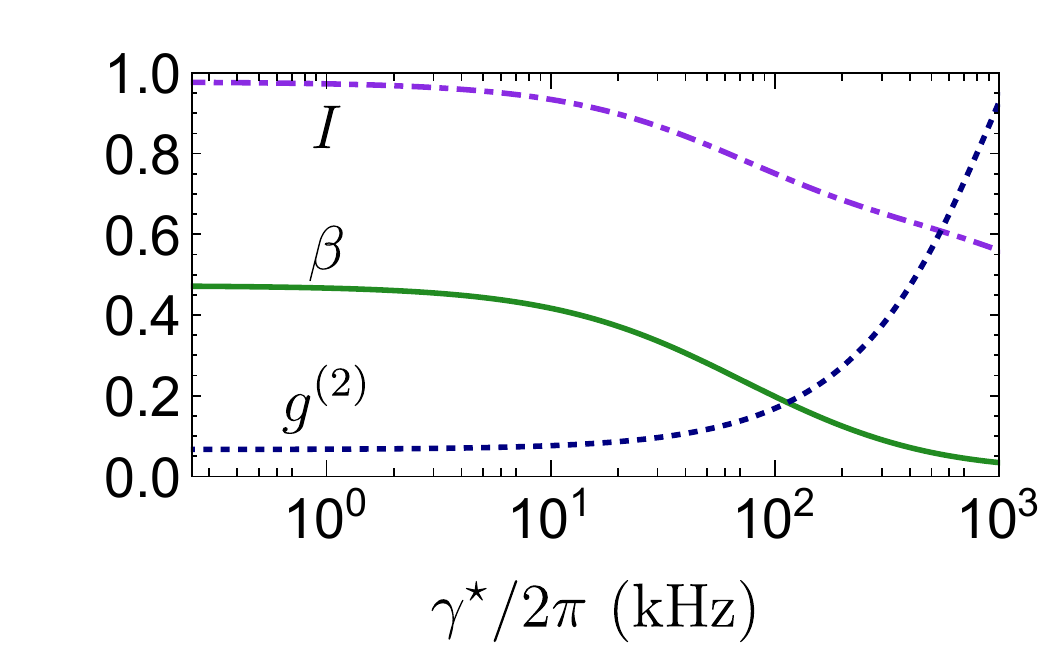}\vspace{-2mm}

\vspace{-5mm}
\begin{flushleft}
\mbox{(e)\hspace{40mm}(f)}
\end{flushleft}\vspace{-5mm}

\hspace{-5mm}
\includegraphics[scale=0.4]{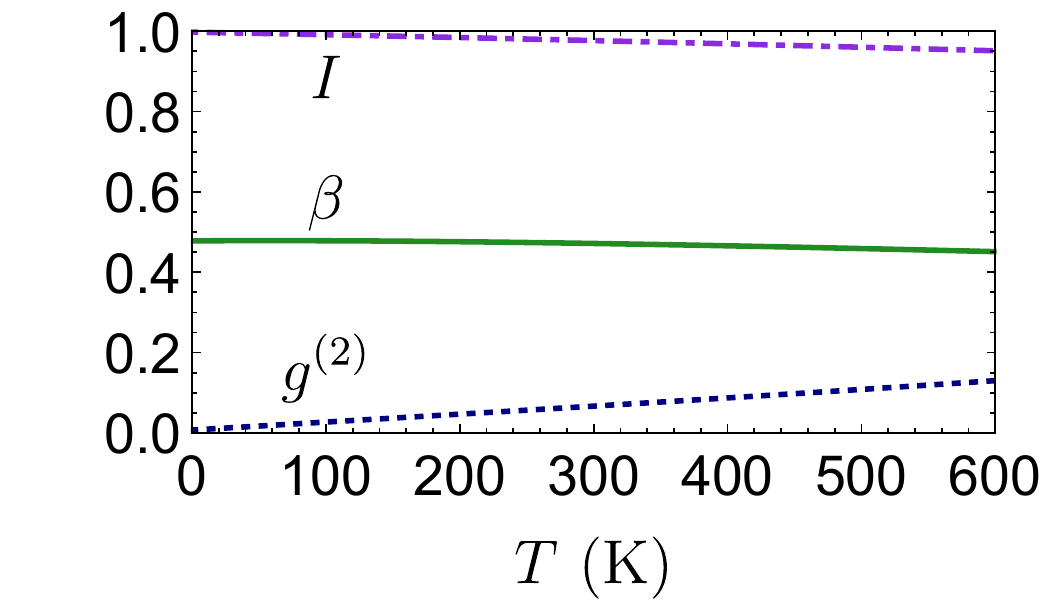}\hspace{-4mm}
\includegraphics[scale=0.4]{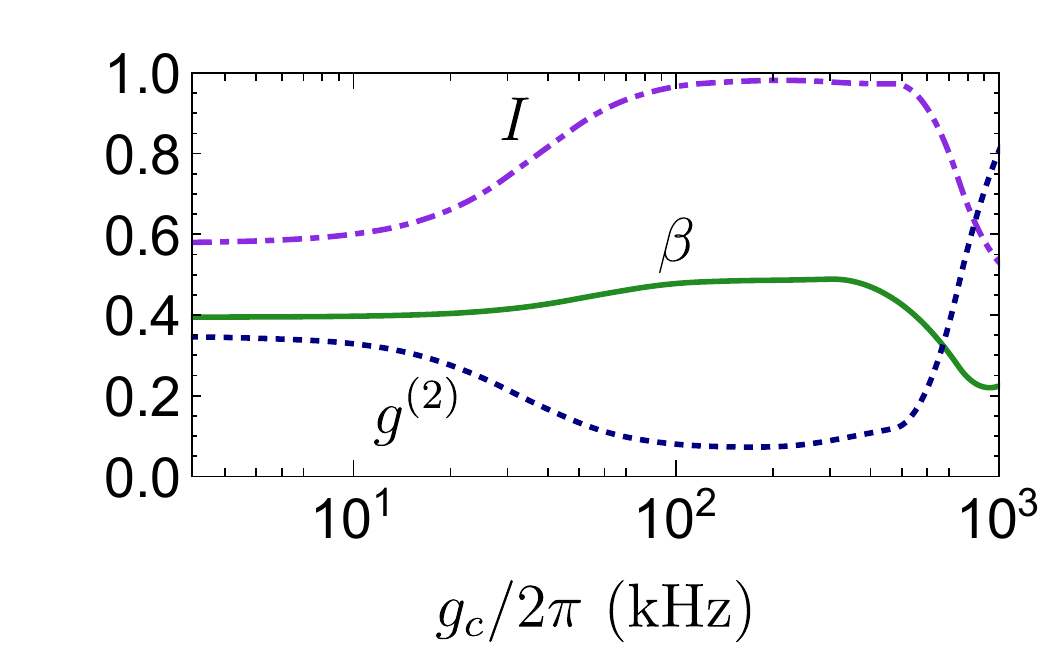}\vspace{-2mm}
\caption{(a) Indistinguishability $I$, brightness $\beta$, and purity $g^{(2)}$, as a function of gate end time $t_u$ for $t_l=0$. (b) Improvement in the indistinguishability at the cost of brightness by decreasing total gate time $t_u^\prime-t_l$ for fixed $t_u^\prime=22\hspace{1mm}\mu$s. (c) Improvement in indistinguishability and purity by increasing mechanical quality factor. Single-photon quality as a function of (d) spin-dephasing rate, (e) temperature, and (f) cooling mode optomechanical coupling rate. Parameters used for all plots: T=300 K, $Q_\text{m}=5\times 10^8$, $f_\text{m}=3$ MHz, $\delta=2\pi\times1$ MHz, $\lambda=g_\text{sp}=2\pi\times100$ kHz, $\kappa_\text{sp}=2\Omega=2\pi\times20$ kHz, $\kappa_\text{c} =4g_\text{c}=2\pi\times 600$ kHz, $\gamma^\star=2\pi\times0.2$ kHz \cite{T2NV}, $t_l=10\hspace{1mm}\mu$s, and $t_u=22\hspace{1mm}\mu$s unless otherwise indicated.}
\label{fig2}
\end{figure}

We find that indistinguishability and purity are optimized around $t_u=\pi/R$ (Fig. \ref{fig2}a). In addition, decreasing the total gate time increases indistinguishability at the cost of brightness, but it cannot arbitrarily improve purity as this is limited by the constant thermal emission rate (Fig. \ref{fig2}b); however, increasing mechanical quality factor improves both indistinguishability and purity (Fig. \ref{fig2}c). Note that with a higher mechanical quality factor, the gate time can be increased to also improve the brightness. We also find that the source is robust for spin-dephasing rates up to $\gamma^\star=\Omega=2\pi\times 10$ kHz (Fig. \ref{fig2}d). Moreover, because the thermal noise scales as $\Gamma_\text{th}\propto T$, the source is also robust against temperature (Fig. \ref{fig2}e). This is made possible by the efficient cooling of the mechanical oscillator. For this, $\kappa_\text{c}$ is chosen to remain in the sideband-resolved regime ($\kappa_\text{c}<\omega_m$) while not being so small as to cause normal mode splitting of the optomechanical states ($g_\text{c}<2\kappa_\text{c}$); $g_\text{c}$ must also be large enough to provide efficient cooling ($4g_\text{c}^2>n_\text{th}\gamma_\text{m}(2n_\text{th}\gamma_\text{m}+\kappa_\text{c})$), leaving a range for high performance (Fig. \ref{fig2}f).

The three crucial requirements for our proposal are ultra-high mechanical quality factor oscillators at MHz frequencies, large spin-mechanics coupling, and optical cavities with a narrow bandwidth. 
Mechanical quality factors as high as $10^{8}$ were reported for SiN trampoline membranes \cite{OM roomT1,OM roomT2,OM roomT3}. Recently, $10^{9}$ was reported by Ghadimi {\it et al.} for nanobeams of the same material \cite{Ghadimi17}. The principles for engineering high quality factors are dissipation dilution \cite{Verbridge06} and strain engineering \cite{OM roomT3, Ghadimi17}. It was shown in \cite{Ghadimi17} that a combination of these two methods for a mechanical oscillator with intrinsic quality factor $Q_{0}$, frequency $f_\text{m}$, thickness $h$, leads to the following scaling $Q_\text{m}\approx Q_{0}\sigma^{2}/3E\rho h^{2} f_\text{m}^{2}$, where $E$ is the Young's module and $\rho$ is the material density. For $f_\text{m}=2$ MHz and the parameters as in \cite{OM roomT1}, one could get quality factors as high as $10^{10}$. Further improvement in $Q_m$ is possible by enhancing $Q_0$ by surface conditioning \cite{Villanueva14}.

For the parameters given in the caption of figure \ref{fig2}, we estimate $I=0.98$, purity $g^{(2)}=0.07$, and brightness $\beta=0.48$. For $f_\text{m}=2$ MHz and $Q_\text{m}=10^{10}$, we estimate it is possible to achieve $I=0.991$, $g^{(2)}=0.0077$, and $\beta=0.70$ using $g_\text{sp}=\lambda=2\pi\times 50$ kHz, $\delta=2\pi\times0.6$ MHz, $\kappa_\text{sp}=2\Omega$, $\kappa_\text{c}=2g_\text{c}=2\pi\times60$ kHz, $t_l=20\hspace{1mm}\mu$s, and $t_u=65\hspace{1mm}\mu$s. Notably, a larger $Q_\text{m}$ reduces the thermal noise and allows for high performance with a slower interaction rate, reducing the required spin-mechanics coupling.

To achieve the spin-mechanics coupling, a magnetic field gradient of $G_\text{m}\sim\! 10^{8}$ T/m is needed for an oscillator with $\sim\!0.1$ ng effective mass. A gradient of $10^{7}$ T/m has already been reported at 25 nm from a magnetic tip \cite{Gm}. Engineering suitable oscillators with $\sim$pg effective masses \cite{Ghadimi17} would reduce the required gradient by one order of magnitude. In addition, recent advances in diamond surface treatment allow stable NV centers to be implanted 5--10 nm below the surface with significantly less degradation to their spin coherence times \cite{zhang2019,deleon}. The closer proximity can increase the gradient by two or three orders of magnitude, allowing a sufficient spin-mechanics coupling. For example, Ref. \cite{deleon} reported an NV center at 10 nm with $\gamma^\star=2\pi\times 5$ kHz, which is within the decoherence tolerance of our proposal. A large magnetic field gradient might shorten the spin coherence time due to magnetic domain noise of the tip material \cite{NVM2}. This could be avoided by using a single-domain rare earth nano-magnet \cite{NVM2, rare}.

For the optomechanical cavity, we consider a $60$ cm long Fabry-P\'{e}rot cavity with a finesse of 12000 ($\kappa_\text{sp}=2\pi\times20$ kHz) driven by a $\sim\!30\hspace{1mm}\mu$W laser. Such a high finesse cavity could be realized by patterning a 2D photonic crystal on the SiN membrane \cite{OM roomT1}. The tip could be located at a node of the optical field to minimize any disturbance to the cavity mode. The reflectivity of the photonic crystal pattern will vary for different spectral modes of the cavity. Thus a more broadband mode ($\kappa_c=2\pi\times 600$ kHz) driven by a \mbox{$\sim\!1$ mW} laser could be used for efficient cooling. The cavity length could be reduced by improving the finesse using a three-mirror design \cite{OM roomT1} or by speeding up the interface. The latter could be achieved using a stimulated adiabatic \cite{vitanov2017} or superadiabatic \cite{zhou2017} Raman scheme by shaping the microwave dressing ($\Omega_\text{q}$) and optomechanical ($g_\text{sp}$) rates.



To summarize, the design we envision would combine a SiN photonic crystal trampoline membrane \cite{OM roomT1} with strain-engineered arms \cite{Ghadimi17} and a single-domain rare earth magnetic tip \cite{NVM2, rare} sitting $\sim 10$ nm away from a shallow-implanted NV center in surface-treated diamond \cite{deleon}. This proposal is adaptable to other configurations. For example, other oscillator geometries could also be successful and more compact. Also, there could exist other defects that have good room-temperature electronic spin coherence but poor optical properties.

In addition to being a single-photon source, the spin-optomechanical interface could generate spin-photon entanglement by first preparing the NV center in \mbox{$(\ket{e}+\ket{0})/\sqrt{2}$}. The emission of a telecom photon from $\ket{e}$ would then create the spin-photon entangled state $(\ket{d}\ket{1}+\ket{0}\ket{0})/\sqrt{2}$. Two spatially-separated devices could then be entangled using the Barrett-Kok procedure \cite{Barrett}. Interestingly, for long-distance quantum networks, the repetition rate is limited by ms communication times \cite{SimonPRL} and so a slow interface is not an issue.




In conclusion, given the recent progress in achieving ultra-high mechanical quality factors, ultra-narrow optomechanical cavity bandwidths, and shallow-implanted NV centers, the realization of room-temperature spin-photon interfaces based on spin-optomechanical transduction is within reach. Together with proposals for non-cryogenic NV-based quantum information processors \cite{Dolde,Yao}, this component could be the missing piece needed to develop a quantum internet that can operate at room-temperature.

\emph{Acknowledgments.} We acknowledge useful discussions with A. H. Ghadimi, S. Goswami, J. P. Hadden, S. Kolkowitz, D. Lake, N. Lauk, H. Ollivier, and C. \O stfeldt. This work was supported by the Natural Sciences and Engineering Research Council of Canada (NSERC) and by Alberta Innovates.


\begin{thebibliography}{}

\bibitem{QKD} N. Gisin, G. Ribordy, W. Tittel, H. Zbinden, Rev. Mod. Phys. {\bf 74}, 145  (2002).

\bibitem{Blind} S. Barz, E. Kashefi, A. Broadbent, J. F. Fitzsimons, A. Zeilinger, P. Walther, Science {\bf 335}, 303 (2012).

\bibitem{Private} M. Jakobi, C. Simon, N. Gisin, J. D. Bancal, C. Branciard, N. Walenta, and H. Zbinden, Phys. Rev. A {\bf 83}, 022301 (2011).

\bibitem{clock} P. Komar, E. M. Kessler, M. Bishof, L. Jiang, A. S. Sørensen, J. Ye, M. D. Lukin, Nat. Phys. \textbf{10}, 582 (2014).

\bibitem{Telescope} D. Gottesman, T. Jennewein, and S. Croke, Phys. Rev. Lett. \textbf{109}, 070503  (2012).

\bibitem{Rideout} D. Rideout {\it et al}, Class. Quant. Grav. {\bf 29}, 224011 (2012).

\bibitem{SimonNP} C. Simon, Nat. Photon. \textbf{11}, 678 (2017).

\bibitem{T2NV} G. Balasubramanian {\it et al}, Nat. Mater. \textbf{8}, 383 (2009).

\bibitem{NuclearT2} P. C. Maurer {\it et al}, Science. \textbf{336},1283 (2012).

\bibitem{Bernien} H. Bernien {\it et al}, Nature \textbf{497}, 86 (2013).

\bibitem{Hensen} B. Hensen {\it et al}, Nature \textbf{526}, 682 (2015).

\bibitem{Fu} K. M. C. Fu, C. Santori, P. E. Barclay, L. J. Rogers, N. B. Manson, and R. G. Beausoleil, Phys. Rev. Lett. \textbf{103}, 256404 (2009).

\bibitem{SangouardRMP} N. Sangouard, C. Simon, H. de Riedmatten, N. Gisin,  Rev. Mod. Phys. \textbf{83}, 33 (2011).

\bibitem{Newton} B. L. Green {\it et al}, Phys. Rev. Lett. \textbf{119}, 096402 (2017).


\bibitem{De Leon} B. C Rose {\it et al}, Science \textbf{361}, 60 (2018).


\bibitem{Wein} S. Wein, N. Lauk, R. Ghobadi, C. Simon, Phys. Rev. B \textbf{97}, 205418 (2018).

\bibitem{RMPom} M. Aspelmeyer, T. J. Kippenberg, and F. Marquardt, Rev. Mod. Phys. \textbf{86}, 1391 (2014).

\bibitem{NVM1} O. Arcizet, V. Jacques, A. Siria, P. Poncharal, P. Vincent and S. Seidelin, Nat. Phys, \textbf{7}, 879 (2011).

\bibitem{NVM2}S. Kolkowitz, A. C. Bleszynski Jayich, Q. Unterreithmeier, S. D. Bennett, P. Rabl, J. G. E. Harris, and M. D. Lukin, Science \textbf{335}, 6076 (2012).

\bibitem{Stannigel12} K. Stannigel, P. Komar, S. J. M. Habraken, S. D. Bennett, M. D. Lukin, P. Zoller, P. Rabl, Phys. Rev. Lett. \textbf{109}, 013603 (2012).

\bibitem{Bagci14} T. Bagci {\it et al}, Nature \textbf{507}, 81 (2014).

\bibitem{Lehnert} R. W. Andrews, R. W. Peterson, T. P. Purdy, K. Cicak, R. W. Simmonds, C. A. Regal, K. W. Lehnert, Nat. Phys {\bf 10}, 321 (2014).

\bibitem{Riedinger16} R. Riedinger, S. Hong, R. A. Norte, J. A. Slater, J. Shang, A. G. Krause, V. Anant, M. Aspelmeyer, S. Gr\"{o}blacher, Nature \textbf{530}, 313 (2016).

\bibitem{Verhagen} E. Verhagen, S. Deleglise,	S. Weis, A. Schliesser, and T. J. Kippenberg, Nature \textbf{482}, 63  (2012).

\bibitem{Wollman} E. E. Wollman, C. U. Lei, A. J. Weinstein, J. Suh, A. Kronwald, F. Marquardt, A. A. Clerk, K. C. Schwab, Science {\bf 349}, 952 (2015).

\bibitem{Galland} C. Galland, N. Sangouard, N. Piro, N. Gisin, and T.J. Kippenberg, Phys. Rev. Lett. {\bf 112}, 143602 (2014).

\bibitem{Hong17} S. Hong, R. Riedinger, I. Marinkovic, A. Wallucks, S. G. Hofer, R. A. Norte, M. Aspelmeyer, S. Gr\"{o}blacher, Science \textbf{358}, 203 (2017).

\bibitem{purdy} T. P. Purdy, K. E. Grutter, K. Srinivasan, and J. M. Taylor, arXiv preprint arXiv:1605.05664 (2016).

\bibitem{sudhir} V. Sudhir, R. Schilling, S. A. Fedorov, H. Schütz, D. J. Wilson, and T. J. Kippenberg, Phys. Rev. X {\bf 7}, 031055 (2017).

\bibitem{OM roomT1} R. A. Norte, J. P. Moura, and S. Gr\"{o}blacher, Phys. Rev. Lett. \textbf{116}, 147202 (2016).

\bibitem{OM roomT2} C. Reinhardt, T. M\"{u}ller, A. Bourassa, and J. C. Sankey, Phys. Rev. X. \textbf{6}, 021001 (2016).

\bibitem{OM roomT3}Y. Tsaturyan, A. Barg, E. S. Polzik and A. Schliesser, Nat. Nanotechnol. \textbf{12}, 776 (2017).

\bibitem{Ghadimi17} A. H. Ghadimi, S. A. Fedorov, N. J. Engelsen, M. J. Bereyhi, R. Schilling, D. J. Wilson, T. J. Kippenberg, arXiv preprint arXiv:1711.06247 (2017).

\bibitem{Indistin1} A. Kiraz, M. Atat\"{u}re, and A. Imamoglu, Phys. Rev. A \textbf{69}, 032305 (2004).

\bibitem{Indistin2} T. Grange, G. Hornecker, D. Hunger, J-P. Poizat, J-M. G\'{e}rard, P. Senellart, and A. Auff\`{e}ves, Phys. Rev. Lett. \textbf{114}, 193601 (2015).

\bibitem{Rabl09} P. Rabl, P. Cappellaro, M. V. Gurudev Dutt, L. Jiang, J. R. Maze and M. D. Lukin, Phys. Rev. B. \textbf{79}, 041302(R)(2009).

\bibitem{Hybrid15} P.B. Li, Y. C. Liu, S.-Y. Gao, Z. L. Xiang, P. Rabl, Y. F. Xiao, and F.L. Li, Phys. Rev. Applied. \textbf{4}, 044003 (2015).

\bibitem{Li16} P.B. Li, Z. L. Xiang, P. Rabl, and F. Nori, Phys. Rev. Lett. \textbf{117}, 015502 (2016).

\bibitem{Dressed state} D. A. Golter, T. K. Baldwin, and H. Wang, Phys. Rev. Lett. \textbf{113}, 237601 (2014).

\bibitem{Verbridge06} S. S. Verbridge, J.M. Parpia, R. B. Reichenbach, L.M. Bellan, and H. G. Craighead, J. Appl. Phys. \textbf{99}, 124304 (2006).

\bibitem{Villanueva14} L. G. Villanueva and S. Schmid. Phy. Rev. Lett, \textbf{113} 227201, (2014).

\bibitem{vitanov2017} N. V Vitanov, A. A. Rangelov, B. W. Shore, and K. Bergmann
Rev. Mod. Phys. \textbf{89}, 015006 (2017).

\bibitem{zhou2017} B. Zhou \emph{et al}, Nat. Phys. \textbf{13}, 330 (2017).

\bibitem{Gm} H. J. Mamin, M. Poggio, C. L. Degen, and D. Rugar, Nat. Nanotechnol. \textbf{2}, 301 (2007).

\bibitem{deleon} S. Sangtawesin \emph{et al}, arXiv preprint arXiv:1811.00144 (2018).

\bibitem{zhang2019} W. Zhang \emph{et al}, arXiv preprint arXiv:1901.05728 (2019).

\bibitem{rare} B. C. Stipe, H. J. Mamin, T. D. Stowe, T. W. Kenny, and D. Rugar, Phys. Rev. Lett. \textbf{86}, 2874 (2001).











\bibitem{Barrett}S. D. Barrett and P. Kok, Phys. Rev. A \textbf{71}, 060310 (2005).

\bibitem{SimonPRL} C. Simon, H. de Riedmatten, M. Afzelius, N. Sangouard, H. Zbinden, and N. Gisin, Phys. Rev. Lett. {\bf 98}, 190503 (2007).





\bibitem{Dolde} F. Dolde, I. Jakobi, B. Naydenov, N. Zhao, S. Pezzagna, C. Trautmann, J. Meijer, P. Neumann, F. Jelezko, J. Wrachtrup, Nat. Phys. \textbf{9}, 139 (2013).

\bibitem{Yao} N. Y. Yao, L. Jiang, A. V. Gorshkov, P. C. Maurer, G. Giedke, J. I. Cirac, M. D. Lukin, Nat. Commun. \textbf{3} 800 (2012).



\end{thebibliography}
\end{document}